\documentclass[prc,aps]{revtex4}
\usepackage{epsfig,amsfonts,amsmath,amssymb,amscd}

\begin{document}

\title{Stellar neutrino energy loss rates due to $^{24}$Mg suitable
for O+Ne+Mg core simulations}

\author{Jameel-Un Nabi\footnote{author e-mail: jnabi00@gmail.com}}

\affiliation{Faculty of Engineering Sciences, GIK Institute of
Engineering Sciences and Technology, Topi 23640, Swabi, NWFP,
Pakistan\\ Current Address: ICTP, Strada Costiera 11, 34014,
Trieste, Italy}
\begin{abstract}
Neutrino losses from proto-neutron stars play a pivotal role to
decide if these stars would be crushed into black holes or explode
as supernovae. Recent observations of subluminous Type II-P
supernovae (e.g., 2005cs, 2003gd, 1999br, 1997D) were able to
rejuvenate the interest in 8-10 M$_{\odot}$ stars which develop
O+Ne+Mg cores. Simulation results of O+Ne+Mg cores show varying
results in converting the collapse into an explosion. The neutrino
energy loss rates are important input parameters in core collapse
simulations. Proton-neutron quasi-particle random phase
approximation (pn-QRPA) theory has been used for calculation of
neutrino energy loss rates due to $^{24}$Mg in stellar matter. The
rates are presented on a detailed density-temperature grid suitable
for simulation purposes. The calculated neutrino energy loss rates
are enhanced up to more than one order of magnitude compared to the
shell model calculations and favor a lower entropy for the core of
these massive stars. \vskip 0.1in\noindent\textbf{PACS} numbers:
23.40.Bw, 21.60.Jz, 26.30.Jk, 26.50.+x\vskip 0.2in
\end{abstract}
 \maketitle
\section{Introduction}

It is now more than forty years since Colgate $\&$ White [1] and
Arnett [2] presented their classical work on energy transport by
neutrinos and antineutrinos in non-rotating massive stars. Despite
immense technological advancements in recent times, the explosion
mechanism of core-collapse supernovae continues to challenge
astrophysicists throughout the globe. When the iron core becomes
unstable against the gravity force its inner portion undergoes
homologous collapse whereas the outer portion collapses
supersonically. The prompt shock that follows the bounce of the core
stagnates (as it looses energy due to neutrino emission and
endothermic dissociation of heavy nuclei falling through the shock
)leading to a standing shock. Some additional source of energy is
required to revitalize the postshock gas. One interesting mechanism
to revive the shock was the "preheating" mechanism proposed by
Haxton [3]. Haxton suggested that a significant fraction of the
energy carried by the neutrinos could be utilized in preheating iron
nuclei outside the shock front rather than getting lost. This would
lead to a reduction in the energy required from the shock wave to
dissociate the infalling Fe, assisting the shock wave to retain more
of its strength as it propagates through the iron core and thus
increasing the chances of transforming the collapse into an
explosion. However, the preheating mechanism continues to be debated
and the part played by neutrinos in this scenario is far from being
completely understood. Later, Bruenn and Haxton [4] worked on models
simulating weak and strong shock cases and found out that in neither
case is the energy transferred to the matter by neutrino-nucleus
absorption significant in terms of preheating the infalling
iron-like material. More recently, Langanke and collaborators [5]
had some observations on the models of Bruenn and Haxton [4] and
reported much larger preshock heating rates. They concluded that
these rates act for too short a time to lead to consequences for
shock propagation. Despite all this, the preheating mechanism in
fact has been adopted by most coders, who are continually putting in
more microphysics to make the simulations more realistic and there
continues to be interest in this problem [6].

During the late stages of evolution, a star mainly looses energy
through neutrinos and this process is fairly independent of the
mass of the star. White dwarfs and supernovae, both have cooling
rates largely dominated by neutrino production. Neutrinos are
crucial to the life and afterlife of a supernova. It is natural to
consider neutrino heating as a probable mechanism for shock
revival as they dominate the energetics and dynamics of the
post-bounce evolution. Having only weak interactions, they are
nature's most efficient means of cooling, transporting along around
99$\%$ of the released gravitational energy. As such it is no
wonder that astrophysicists find it a formidable task to simulate
a 1$\%$ effect of explosion.

White dwarfs located in a binary system may end their lives in two
possible ways. They may accrete from a companion and achieve the
Chandrasekhar mass thereby triggering a thermonuclear runaway of
the object and ultimately exploding as a Type Ia supernova. In
this case no remnant is left behind. Alternatively these massive
white dwarfs, for sufficiently high mass accretion rates, may
allow the formation of O+Ne+Mg cores and due to the high
prevailing central density (beyond $10^{10} g cm^{-3}$) experience
rapid electron capture that lead to the collapse of the core. This
is termed as accretion-induced collapse (AIC). The end product is
a neutron star (in some double-degenerate scenario, two white
dwarfs in a short-period binary system may eventually coalesce to
form a massive white dwarf that exceeds the Chandrasekhar mass
limit and in this case transition to a black hole is possible if
the total proto-neutron star mass exceeds the general-relativistic
limit for gravitational stability [7]). The AIC of white dwarfs
represents one instance where a neutrino mechanism leads
undoubtedly to a successful, albeit weak, explosion [8]. The
authors in Ref. [8] argued that the neutrino mechanism can
successfully power explosions of low-mass progenitors and AICs due
to the limited mantle mass and steeply declining accretion rate.
The ultimate fate of these white dwarfs are dependent on many
factors, e.g. temperature of the environment, the mass accretion
rate on the newly formed white dwarf, the mass of each partner
white dwarf [7] and rapid electron capture rates [8]. The
occurrence rate of the AIC of white dwarfs is not determined
reliably, and are not expected to occur more than once per 20 --
50 standard Type Ia events (see for example [7]).

Supernovae are also tipped in as probable sites for the r-process.
The production site of the heavy r-process nuclei is associated with
the accretion-induced collapse of an oxygen-neon-magnesium (O+Ne+Mg)
white dwarf in a binary system (e.g. [9]) or Type-II supernovae from
8--10 $M_{\odot}$ (e.g. [10]). These 8--10 $M_{\odot}$ stars form an
electron-degenerate O+Ne+Mg core that does not undergo further
nuclear burning. A few collapse simulators (e.g. Hillebrandt et al.
[11]) demonstrated that the collapsing O+Ne+Mg core explodes
promptly. Others (e.g. Woosley and Baron [12]) showed that the shock
generated at core bounce stalls rather than leading to a prompt
explosion. Delayed explosion scenario of these cores was also
reported (e.g. Mayle and Wilson [13]). Wheeler and collaborators
[14] have suggested that the exploding O+Ne+Mg core could be a
viable site for the r-process. However the question of whether
O+Ne+Mg cores can explode hydrodynamically continues to be argued.
For example, Guti\'{e}rrez et al. [15] argued that the abundance of
$^{24}$Mg was considerably reduced in updated evolutionary
calculations. However the procedure adopted by them was not fully
consistent as they kept the ratio of oxygen to neon constant while
parameterizing the abundance of $^{24}$Mg. Later Kitaura et al. [16]
presented simulation result of these cores (keeping capture rates on
$^{24}$Mg as a key ingredient) using an improved neutrino transport
treatment. Their outcome was not a prompt but a delayed explosion.

A smaller iron core present at the onset of the core bounce as
well as the smaller gravitational potential of the collapsing
cores of a 8--10 $M_{\odot}$ star do favor a prompt explosion.
What remains certain is that microscopic and reliable electron
capture rates and neutrino energy loss rates are needed for a
careful study of the late stages of the stellar evolution of 8--10
$M_{\odot}$ stars and can contribute effectively in the final
outcome of the simulations of O+Ne+Mg cores on world's fastest
supercomputers.

Electron capture rates and the accompanying neutrino energy loss
rates of matter in nuclear statistical equilibrium under conditions
of high temperatures and densities are of prime importance in
determining the equation of state of exploding stars. An accurate
determination of neutrino emission rates is mandatory in order to
perform a careful analysis of the final branches of star
evolutionary tracks. A change in the cooling rates at the very last
stages of massive star evolution could affect the evolutionary time
scale and the configuration of iron core at the onset of the
supernova explosion.

Compared to shell model calculations by Wildenthal et al. [17],
the proton-neutron quasiparticle random phase approximation theory
(pn-QRPA) gives similar accuracy in the description of light
nuclei including beta-decay rates in $sd$-shell nuclide [18]. Nabi
and Klapdor [19], later, calculated weak interaction rates for 709
nuclei with A = 18 to 100 in stellar matter using the pn-QRPA
theory. These included capture rates, decay rates, neutrino energy
loss rates, probabilities of beta-delayed particle emissions and
energy rate of these particle emissions (see also Refs. [20,21]).
Since then these calculations were further refined with use of
more efficient algorithms, incorporation of latest data from mass
compilations and experimental values, and fine-tuning of model
parameters (e.g. Refs. [22 -- 25]).

The evolution of the stars in the mass range 8-10 M$_{\odot}$
develops central cores which are composed of $^{16}$O, $^{20}$Ne and
$^{24}$Mg. Oda et al. [26] pointed out three different series of
electron capture in the O+Ne+Mg core of the 8-10 M$_{\odot}$ stars
and placed them in the order of low threshold energy as $^{24}$Mg
$\rightarrow$ $^{24}$Na $\rightarrow$ $^{24}$Ne, $^{20}$Ne
$\rightarrow$ $^{20}$F $\rightarrow$ $^{20}$O, and $^{16}$O
$\rightarrow$ $^{16}$N $\rightarrow$ $^{16}$C.  The first series was
regarded most important as the trigger nucleus, $^{24}$Mg, has the
lowest electron capture threshold. Oda and collaborators [26]
actually renounced the last series in their calculations because of
its high threshold energy which did not contribute significantly to
the initiation of the collapse of the O+Ne+Mg core of the 8-10
M$_{\odot}$ stars.

The Gamow-Teller strength distributions in $^{24}$Mg and the
calculation of electron capture rates on $^{24}$Mg using the pn-QRPA
theory for O+Ne+Mg core simulations were presented earlier by Nabi
and Rahman [23]. Here I present for the first time the pn-QRPA
calculations of the neutrino and antineutrino energy loss rates due
to $^{24}$Mg in stellar plasma on a detailed temperature-density
grid suitable for the simulation codes of O+Ne+Mg cores. In order to
further increase the reliability of the calculated neutrino energy
loss rates, I have incorporated all measured excitation energies
(along with their log $ft$ values) with definite spin and/or parity
assignment, by either replacement (when they were within 0.5 MeV of
each other) or manual insertion. The deformation parameter,
$\delta$, is recently argued to be one of the most important
parameters in QRPA calculations [27].  For the case of even-even
nuclei, experimental deformations are available [28] and were
employed in this work. All these steps were taken to ensure the most
reliable calculation of pn-QRPA neutrino and antineutrino energy
loss rates due to $^{24}$Mg in presupernova and supernova
environment.

The paper is written as follows. Section II deals with the brief
formalism of the pn-QRPA calculations. I present some of my
calculated results in Section III. Comparisons with earlier
calculations is also included in this section. I finally conclude in
Section IV and at the end Table II presents the detailed calculation
of neutrino and antineutrino energy loss rates due to $^{24}$Mg
suitable for simulation codes.

\section{Model description}
I used the pn-QRPA theory with a separable interaction to
calculate the neutrino and antineutrino energy losses in stellar
matter. The Hamiltonian of the problem was taken to be of the
form:
\begin{equation}
H^{QRPA} =H^{sp} +V^{pair} +V_{GT}^{ph} +V_{GT}^{pp}.
\end{equation}

For the single-particle Hamiltonian, $H^{sp}$, the single particle
energies as well as wave functions were calculated in the Nilsson
model (which takes into account nuclear deformations). The BCS model
was used to calculate the pairing force, $V^{pair}$. Two types of
proton-neutron residual interactions were incorporated in these
calculations namely the particle-hole, $V_{GT}^{ph}$, and the
particle-particle, $V_{GT}^{pp}$, interaction. The two interactions
(separable in form) were characterized by two interaction constants:
$\chi$ (for particle-hole interaction) and $\kappa$ (for
particle-particle interaction). In this work, the values of $\chi$
and $\kappa$ was taken as 0.001 MeV and 0.05 MeV, respectively. (The
reader is referred to [18,29] for details of optimum selection of
these parameters.) Other parameters required for the calculation of
weak rates are the deformation, the pairing gaps, and the Q-value of
the nuclear reactions. The calculated half-lives depend only weakly
on the values of the pairing gaps [30]. Thus, the traditional choice
of $\Delta _{p} =\Delta _{n} =12/\sqrt{A} (MeV)$ was applied in the
present work. The deformation parameter is as an important parameter
for QRPA calculations as pairing [27]. As such rather than using
deformations from some theoretical mass models (as used in earlier
calculations of pn-QRPA capture rates) the experimentally adopted
value of the deformation parameters for $^{24}$Mg, extracted by
relating the measured energy of the first $2^{+}$ excited state with
the quadrupole deformation, was taken from Raman et al. [28].
Q-values were taken from the recent mass compilation of Audi et al.
[31].

An estimation of uncertainties of the model and theoretical errors
is important to establish the reliability of the calculated rates.
The uncertainties related to the pn-QRPA model was discussed in
detail in Ref. [32]. There the accuracy of the pn-QRPA model was
compared to the experimental data for both $\beta^{+}$ and
$\beta^{-}$ directions (see Table I and Table II of Ref. [32]) and
an estimation of theoretical errors was given. The decay rate of
even-even nuclei generally exhibits more or less a smooth behavior
with respect to the deformation parameter (see Fig.2 of Ref. [18]
and discussions therein). The dependence of the half-life on the
pairing gaps is small and rather smooth (see Fig. 3 of Ref. [18]).
The half-lives are not very sensitive to a change of the
Gamow-Teller interaction strength $\chi$ within a reasonable range
and the relationship is almost linear [18]. On the other hand for
large values of $\kappa$ the lowest eigenvalue becomes complex.
This indicates the "collapse" of the QRPA model and it is crucial
to avoid values of $\kappa$ which lie beyond the collapse. The
redistribution of the $\beta$ strength and the reduction of the
sum of the Gamow-Teller strength lead to a nonlinear dependence of
the calculated half-lives as a function of $\kappa$ [29]. Figs. (7
-- 10) of Ref. [29] show in detail how the observables of the
pn-QRPA calculations vary with changes in $\kappa$. These analysis
do explain that a small and acceptable change in the QRPA
parameters does not have a dramatic dependence on the calculated
rates.

The neutrino energy loss rates can occur through four different
weak-interaction mediated channels: electron and positron
emissions, and, continuum electron and positron captures. The
neutrino energy loss rates were calculated using the formula

\begin{equation}
\lambda ^{^{\nu} } _{ij} =\left[\frac{\ln 2}{D}
\right]\left[f_{ij}^{\nu} (T,\rho ,E_{f} )\right]\left[B(F)_{ij}
+\left({\raise0.7ex\hbox{$ g_{A}  $}\!\mathord{\left/ {\vphantom
{g_{A}  g_{V} }} \right.
\kern-\nulldelimiterspace}\!\lower0.7ex\hbox{$ g_{V}  $}}
\right)^{2} B(GT)_{ij} \right].
\end{equation}
The value of D was taken to be 6295s [33] and the ratio of the
axial vector to the vector coupling constant as -1.254 [34].
$B_{ij}'s$ are the sum of reduced transition probabilities of the
Fermi B(F) and Gamow-Teller transitions B(GT). The $f_{ij}^{\nu}$
are the phase space integrals and are functions of stellar
temperature ($T$), density ($\rho$) and Fermi energy ($E_{f}$) of
the electrons. They are explicitly given by
\begin{equation}
f_{ij}^{\nu} \, =\, \int _{1 }^{w_{m}}w\sqrt{w^{2} -1} (w_{m} \,
 -\, w)^{3} F(\pm Z,w)(1- G_{\mp}) dw,
\end{equation}
and by
\begin{equation}
f_{ij}^{\nu} \, =\, \int _{w_{l} }^{\infty }w\sqrt{w^{2} -1}
(w_{m} \,
 +\, w)^{3} F(\pm Z,w)G_{\mp} dw.
\end{equation}
In above equation $w$ is the total energy of the electron
including its rest mass, $w_{l}$ is the total capture threshold
energy (rest+kinetic) for positron (or electron) capture. F($ \pm$
Z,w) are the Fermi functions and were calculated according to the
procedure adopted by Gove and Martin [35]. G$_{\pm}$ is the
Fermi-Dirac distribution function for positrons (electrons).
\begin{equation}
G_{+} =\left[\exp \left(\frac{E+2+E_{f}
}{kT}\right)+1\right]^{-1},
\end{equation}
\begin{equation}
 G_{-} =\left[\exp \left(\frac{E-E_{f} }{kT}
 \right)+1\right]^{-1},
\end{equation}
here $E$ is the kinetic energy of the electrons and $k$ is the
Boltzmann constant.

For the decay channel Eqt. (3) was used for the calculation of
phase space integrals. Upper signs were used for the case of
electron emissions and lower signs for the case of positron
emissions. Regarding the capture channels, I used Eqt. (4) for the
phase space integrals keeping upper signs for continuum electron
captures and lower signs for continuum positron captures.

The total neutrino energy loss rate per unit time per nucleus is
given by
\begin{equation}
\lambda^{\nu} =\sum _{ij}P_{i} \lambda _{ij}^{\nu},
\end{equation}
where $\lambda_{ij}^{\nu}$ is the sum of the electron capture and
positron decay rates for the transition $i \rightarrow j$ and
$P_{i}$ is the probability of occupation of parent excited states
which follows the normal Boltzmann distribution.

On the other hand the total antineutrino energy loss rate per unit
time per nucleus is given by
\begin{equation}
\lambda^{\bar{\nu}} =\sum _{ij}P_{i} \lambda _{ij}^{\bar{\nu}},
\end{equation}
where $\lambda_{ij}^{\bar{\nu}}$ is the sum of the positron
capture and electron decay rates for the transition $i \rightarrow
j$.

The pn-QRPA theory allows a microscopic state-by-state calculation
of both sums present in Eqts. (7) and (8). In other words the
pn-QRPA theory calculates the GT strength distribution of all
excited states of parent nucleus in a microscopic fashion. This
feature of the pn-QRPA model greatly increases the reliability of
the calculated rates in stellar matter where there exists a finite
probability of occupation of excited states. Further the pn-QRPA
theory can handle any arbitrarily heavy system of nucleons since
the calculation is performed in a luxurious model space of up to 7
major oscillator shells.

Details of the calculations of phase space integrals and reduced
transition probabilities can be found in [20].

\section{Results and comparison}
The Gamow-Teller strength distributions, B(GT$_{\pm}$), for
$^{24}$Mg were calculated using the pn-QRPA theory. Quenching of
the GT strength was taken into account and a standard quenching
factor of 0.77 was used. The calculation was performed for a total
of 132 excited states of $^{24}$Mg up to excitation energies in
the vicinity of 40 MeV (corresponding to first 100 excited states)
in the daughters $^{24}$Na and $^{24}$Al. The total calculated
B(GT$_{+}$) and B(GT$_{-}$) strength came out to be 3.34
(quenched) and Ikeda sum rule was satisfied in the calculations.

In this section I present some of the results of the pn-QRPA
calculated (anti)neutrino energy loss rates due to $^{24}$Mg. The
reported rates are also compared against previous calculations.

Fig.1 and Fig. 2 show four panels depicting the calculated neutrino
and antineutrino energy loss rates, respectively, at selected
temperature and density domain. It is pertinent to mention again
that the neutrino energy loss rates (depicted in Fig.1) contain
contributions due to electron capture \textit{and} positron decay on
$^{24}$Mg whereas the antineutrino energy loss rates (Fig.2) are
calculated due to contributions from positron capture \textit{and}
electron decay on $^{24}$Mg. The upper left panel, in both figures,
shows the energy loss rates in low-density region ($\rho [gcm^{-3}]
=10^{0.5}, 10^{1.5}$ and $10^{2.5}$), the upper right in medium-low
density region ($\rho [gcm^{-3}] =10^{3.5}, 10^{4.5}$ and
$10^{5.5}$), the lower left in medium-high density region ($\rho
[gcm^{-3}] =10^{6.5}, 10^{7.5}$ and $10^{8.5}$) and finally the
lower right panel depicts the calculated rates in high density
region ($\rho [gcm^{-3}] =10^{9.5}, 10^{10.5}$ and $10^{11}$). The
(anti)neutrino energy loss rates are given in logarithmic scales (to
base 10) in units of $MeV. s^{-1}$. In the figures T$_{9}$ gives the
stellar temperature in units of $10^{9}$ K. One should note the
order of magnitude differences in neutrino energy loss rates as the
stellar temperature increases (Fig. 1). It can be seen from this
figure that in the low density region the energy loss rates, as a
function of stellar temperatures, are more or less superimposed on
one another. This means that there is no appreciable change in the
neutrino energy loss rates when increasing the density by an order
of magnitude. There is a sharp exponential increase in the neutrino
energy loss rates for the low, medium-low and medium-high density
regions as the stellar temperature increases to T$_{9}$ =5. Beyond
this temperature the slope of the rates reduces drastically. One
also observes that the neutrino energy loss rates are almost the
same for the densities in the range $(10-10^{6})g/cm^{3}$ as a
function of stellar temperature. For a given temperature the
neutrino energy loss rates increase monotonically with increasing
densities.

As far as the antineutrino energy loss rates are concerned (Fig.
2) one again notes that the rates increase very sharply as T$_{9}$
approaches 5. The rates are again almost superimposed on one
another as a function of stellar densities. However as the stellar
matter moves from the medium high density region to high density
region these rates start to 'peel off' from one another. The
antineutrino production rates are dominated by the capture of
positrons due to $^{24}$Mg with a relatively smaller contribution
coming from the beta-decay of $^{24}$Mg (see Eqt. (8)).

The current calculation was also compared with two earlier
calculations of neutrino and antineutrino energy loss rates due to
$^{24}$Mg. Fuller, Fowler and Newman [36] (hereafter FFN) compiled
the experimental data and calculated neutrino and antineutrino
energy loss rates (besides other weak-interaction mediated rates)
for the nuclei in the mass range A = 21-60 for an extended grid of
temperature and density. The GT strength and excitation energies
were calculated using a zero-order shell model. For the discrete
transitions, for which the $ft$ values were not available, FFN
took log $ft$ = 5.0. Later Oda et al. [26] did an extensive
calculation of stellar weak interaction rates of $sd$-shell nuclei
in full (sd)$^{n}$-shell model space. They used the effective
interaction of Wildenthal [17] and the available experimental
compilations for their calculations.

Fig.3 (Fig.4) depicts the comparison of the pn-QRPA neutrino
(antineutrino) energy loss rates with those calculated using shell
model (OHMTS) [26] and those by FFN [36]. The ordinate represents
the log (to base 10) of the calculated energy loss rates in units
of MeV/s. The left panel depicts the different calculated energy
loss rates at density $10^{3} gcm^{-3}$, the middle panel at
density $10^{7} gcm^{-3}$ and the right panel at density $10^{11}
gcm^{-3}$. The shell model rates are usually in good comparison
with the corresponding FFN rates. The calculated pn-QRPA rates are
enhanced. Table I shows the ratio of the calculated neutrino
energy loss rates to those of Oda et al. [26] and FFN [36]
calculations. In the table $R_{\nu}(QRPA/OHMTS)$ denotes the ratio
of the reported neutrino energy loss rates to those calculated
using the shell model whereas $R_{\nu}(QRPA/FFN)$ gives the
corresponding ratio for the FFN calculations. The reported rates
are more than a factor 15 enhanced as compared to the shell model
rates in the presupernova conditions. The shell model rates are
slightly enhanced at T$_{9}$ = 30 (at high densities the pn-QRPA
neutrino energy loss rates again surpass the shell model rates).
The collapse simulators should take note of this enhanced neutrino
energy loss rates which favor cooler cores with a lower entropy.
It is important to note from Table I that, in comparison to
previous calculations, the QRPA rates are most enhanced around
T$_{9} \sim 1.5 - 2$ (i.e. during the presupernova evolution of
O+Ne+Mg cores). The ratio then starts decreasing with increasing
T$_{9}$. The enhancement of the QRPA rates around T$_{9} \sim 1.5
- 2$ may be traced back to the enhanced electron capture rates on
$^{24}$Mg [37] leading to an enhanced production of non-thermal
neutrinos around these temperatures. It may also be seen from
Table I that at high temperatures (T$_{9} \sim 30$) the ratios are
in good agreement with previous calculations.

The story is different for the comparison of antineutrino energy
loss rates (Fig.4). This time the shell model rates and FFN rates
are much more enhanced compared to pn-QRPA calculations.
Nevertheless the antineutrino energy loss rates are very small
numbers and can change by orders of magnitude by a mere change of
0.5 MeV, or less, in parent or daughter excitation energies and are
more indicative of the uncertainties present in the calculation of
the excitation energies.

Table II finally presents the calculated neutrino and antineutrino
energy loss rates due to $^{24}$Mg on a detailed temperature-density
grid suitable for simulation purposes. Here Column 1 shows the
stellar density in logarithmic scales to base 10 (in units of
$gcm^{-3}$), Column 2 the stellar temperature in units of $10^{9}$
 $K$. Stated also are the values of the Fermi energy of electrons in
units of $MeV$ in Column 3 whereas Column 4 and Column 5 display the
corresponding calculated neutrino and antineutrino energy loss
rates, respectively, in logarithmic scales to base 10 (in units of
$MeV.sec^{-1}$). The ASCII file of Table II is also available and
can be received from the author upon request.

\section{Conclusions}
There exists a wide variety of discrepant results in the supernova
simulations of O+Ne+Mg cores including prompt explosions, delayed
explosions and no explosions. Varying results of the simulations of
O+Ne+Mg cores ask for a more careful analysis of the precollapse
evolution. A large number of input parameters are required for these
simulations to run and a reliable calculation of these parameters
can certainly reduce the large uncertainties involved. Neutrino
energy loss rates are one of the key input nuclear physics
parameters responsible for cooling the stellar cores and thereby
reducing its entropy. The pn-QRPA calculations of the neutrino
energy loss rates due to $^{24}$Mg are presented here on a detailed
temperature-density grid suitable for simulation codes. The pn-QRPA
calculations gives similar accuracy in reproducing weak interaction
mediated rates in $sd$-shell nuclide [18] and is employed here using
maximum possible experimental incorporations to further increase the
reliability of the calculated numbers.

Recent simulations of O+Ne+Mg cores [15,16] employ the weak
interaction rates of shell model [38,26], respectively. The
reported neutrino energy loss rates are up to a factor of 15
enhanced as compared to shell model results. These enhanced
numbers favor cooler cores with lower entropies. Can the enhanced
reported neutrino energy loss rates lead to any change in the
outcome of simulation results on O+Ne+Mg cores? Let us recall that
in the simulations neutrino heating is a favored mechanism for
revival of the stalled shock wave.  The reported neutrino energy
loss rates due to $^{24}$Mg and other nuclei of astrophysical
importance (see e.g. Ref. [32]) employed preferably in a
multi-dimensional model, with neutrino transport included
consistently throughout the entire mass, accounting also for a
complete treatment of multidimensional convection and burning
phases might lead  to some interesting outcome and answers.

\textbf{ACKNOWLEDGMENTS}

The author would like to acknowledge the local hospitality provided
by the Abdus Salam ICTP, Trieste, where part of this project was
completed.

\newpage
\vspace{1.5in} \textbf{Table I:} Ratios of QRPA calculated
neutrino energy loss rates (due to $^{24}$Mg) to those of OHMTS
[26] and FFN [36]. First column gives the grid point, containing
the log (to base 10) of stellar density and temperature in units
of gcm$^{-3}$ and 10$^{9}$ K, respectively, at which these rates
were calculated.
\begin{center}
\begin{tabular}{ccc} \\ \hline
(log($\rho Y_{e}$), T$_{9}$ ) & $R_{\nu}(QRPA/OHMTS)$ &
$R_{\nu}(QRPA/FFN)$\\\hline
(1,1) &1.24E+01 & 1.13E+01\\
(1,1.5) &1.48E+01 & 1.14E+01\\
(1,2) &1.48E+01 & 9.84E+00\\
(1,3) &1.13E+01 & 7.01E+00\\
(1,5) &5.51E+00 & 3.91E+00\\
(1,10) &1.69E+00 & 1.95E+00\\
(1,30) &7.14E-01 & 1.61E+00\\
(3,1) &1.24E+01 & 1.13E+01\\
(3,1.5) &1.48E+01 & 1.14E+01\\
(3,2) &1.48E+01 & 9.84E+00\\
(3,3) &1.13E+01 & 7.01E+00\\
(3,5) &5.52E+00 & 3.92E+00\\
(3,10) &1.69E+00 & 1.95E+00\\
(3,30) &7.16E-01 & 1.61E+00\\
(7,1) &6.71E+00 & 5.98E+00\\
(7,1.5) &1.27E+01 & 9.68E+00\\
(7,2) &1.41E+01 & 9.35E+00\\
(7,3) &1.14E+01 & 7.05E+00\\
(7,5) &5.85E+00 & 4.08E+00\\
(7,10) &1.71E+00 & 1.96E+00\\
(7,30) &7.16E-01 & 1.62E+00\\
(11,1) &3.86E+00 & 3.80E+00\\
(11,1.5) &3.87E+00 & 3.79E+00\\
(11,2) &3.86E+00 & 3.78E+00\\
(11,3) &3.84E+00 & 3.67E+00\\
(11,5) &3.60E+00 & 3.24E+00\\
(11,10) &2.90E+00 & 2.20E+00\\
(11,30) &1.28E+00 & 1.77E+00\\
\hline
\end{tabular}
\end{center}

\newpage
\textbf{Table II:} Calculated neutrino and antineutrino energy loss
rates due to $^{24}$Mg for different selected densities and
temperatures in stellar matter. log($\rho Y_{e}$) has units of $g
cm^{-3}$, where $\rho$ is the baryon density and $Y_{e}$ is the
ratio of the electron number to the baryon number. Temperatures
($T_{9}$) are measured in $10^{9}$ K. $E_{f}$ is the total Fermi
energy of electrons including the rest mass ($MeV$). $\lambda^{\nu}$
are the total neutrino energy loss rates $(MeV s^{-1})$ due to
$\beta^{+}$ decay and electron capture on $^{24}$Mg .
$\lambda^{\bar{\nu}}$ are the total antineutrino energy loss rates
$(MeV s^{-1})$ due to $\beta^{-}$ decay and positron capture on
$^{24}$Mg. The calculated rates are tabulated in logarithmic (to
base 10) scale.
In the table, -100 means that the rate is smaller than 10$^{-100}$. \\
\tiny{\begin{center}
\begin{tabular}{ccccc|ccccc|ccccc} \\ \hline
$log\rho Y_{e}$& $T_{9}$ & $E_{f}$ & $\lambda^{\nu}$ &
$\lambda^{\bar{\nu}}$ & $log\rho Y_{e}$& $T_{9}$ & $E_{f}$ &
$\lambda^{\nu}$ & $\lambda^{\bar{\nu}}$ & $log\rho Y_{e}$& $T_{9}$
& $E_{f}$ & $\lambda^{\nu}$ & $\lambda^{\bar{\nu}}$   \\\hline

0.5& 0.5& 0.065&   -66.886& -100&    1&   8.5& 0&   -3.497&  -7.357&  2&   4.5& 0&   -8.258&  -15.941\\
0.5& 1&   0&   -35.745& -73.995& 1&   9&   0&   -3.156&  -6.785&  2&   5&   0&   -7.319&  -14.173\\
0.5& 1.5& 0&   -24.62&  -49.582& 1&  9.5& 0&   -2.843&  -6.267&  2&   5.5& 0&   -6.529&  -12.71\\
0.5& 2&   0&   -18.795& -37.182& 1&   10&  0&   -2.555&  -5.795&  2&   6&  0&   -5.851&  -11.477\\
0.5& 2.5& 0&   -15.167& -29.65&  1&   15&  0&   -0.505&  -2.612&  2&   6.5& 0&   -5.263&  -10.422\\
0.5& 3&   0&   -12.667& -24.573& 1&   20&  0&   0.773&   -0.8&    2&   7&   0&   -4.745&  -9.506\\
0.5& 3.5& 0&  -10.825& -20.907& 1&   25&  0&   1.697&   0.427&   2&   7.5& 0&   -4.284&  -8.703\\
0.5& 4&   0&   -9.401&  -18.128& 1&   30&  0&   2.416&   1.341&   2&   8&   0&   -3.871&  -7.992\\
0.5& 4.5& 0&   -8.26&   -15.942& 1.5& 0.5& 0.162&   -65.908& -100&    2&   8.5& 0&   -3.496&  -7.356\\
0.5& 5&   0&   -7.321&  -14.175& 1.5& 1&   0.002&   -35.737& -74.001& 2&   9&   0&   -3.155&  -6.784\\
0.5& 5.5& 0&   -6.531&  -12.712& 1.5& 1.5& 0&   -24.618& -49.582& 2&   9.5& 0&   -2.842&  -6.266\\
0.5& 6&   0&   -5.854&  -11.479& 1.5& 2&   0&  -18.794& -37.182& 2&   10&  0&   -2.554&  -5.794\\
0.5& 6.5& 0&   -5.265&  -10.424& 1.5& 2.5& 0&   -15.166& -29.65&  2&   15& 0&   -0.504&  -2.611\\
0.5& 7&   0&   -4.747&  -9.508&  1.5& 3&   0&   -12.666& -24.572& 2&   20&  0&   0.774&   -0.799\\
0.5& 7.5& 0&   -4.287&  -8.705&  1.5& 3.5& 0&   -10.824& -20.906& 2&   25&  0&   1.698&   0.429\\
0.5& 8&   0&   -3.873&  -7.994&  1.5& 4&   0&   -9.399&  -18.126& 2&   30&  0&   2.417&   1.343\\
0.5& 8.5& 0&   -3.499&  -7.359&  1.5& 4.5& 0&   -8.259&  -15.941& 2.5& 0.5& 0.261&   -64.907& -100\\
0.5& 9&   0&   -3.158&  -6.787&  1.5& 5&   0&   -7.319&  -14.173& 2.5& 1&   0.015&   -35.669& -74.07\\
0.5& 9.5& 0&   -2.845&  -6.269&  1.5& 5.5& 0&   -6.529&  -12.711& 2.5& 1.5& 0.002&   -24.613& -49.586\\
0.5& 10&  0&   -2.557&  -5.797&  1.5& 6&   0&   -5.852&  -11.478& 2.5& 2&   0&   -18.793& -37.183\\
0.5& 15&  0&   -0.507&  -2.615&  1.5& 6.5& 0&   -5.263&  -10.422& 2.5& 2.5& 0&   -15.165&-29.65\\
0.5& 20&  0&   0.77&    -0.803&  1.5& 7&   0&   -4.745&  -9.506&  2.5& 3&   0&   -12.666& -24.572\\
0.5& 25&  0&   1.694&   0.424&   1.5& 7.5& 0&   -4.284&  -8.703&  2.5& 3.5& 0&   -10.823& -20.906\\
0.5& 30&  0&   2.412&   1.337&   1.5& 8&   0&   -3.871&  -7.992&  2.5& 4&   0&   -9.399&  -18.126\\
1&   0.5& 0.113&   -66.406& -100&    1.5& 8.5& 0&   -3.496&  -7.356&  2.5& 4.5& 0&   -8.258&  -15.941\\
1&   1&   0&   -35.742& -73.996& 1.5& 9&   0&   -3.155&  -6.784&  2.5& 5&   0&   -7.319& -14.173\\
1&   1.5& 0&   -24.619& -49.582& 1.5& 9.5& 0&   -2.843&  -6.266&  2.5& 5.5& 0&   -6.528&  -12.71\\
1&   2&   0&   -18.794& -37.182& 1.5& 10&  0&   -2.554&  -5.794&  2.5& 6&   0&   -5.851&  -11.477\\
1&   2.5& 0&   -15.166& -29.65&  1.5& 15&  0&   -0.504&  -2.611&  2.5& 6.5& 0&  -5.263&  -10.422\\
1&   3&   0&   -12.666& -24.572& 1.5& 20&  0&   0.774&   -0.799&  2.5& 7&   0&   -4.745&  -9.506\\
1&   3.5& 0&   -10.824& -20.906& 1.5& 25&  0&   1.698&   0.428&   2.5& 7.5& 0&   -4.284&  -8.703\\
1&   4&   0&   -9.4&    -18.127& 1.5& 30&  0&   2.416&   1.342&   2.5& 8 &  0&   -3.87&   -7.991\\
1&   4.5& 0&   -8.259&  -15.941& 2&   0.5& 0.212&   -65.408& -100&    2.5& 8.5& 0&   -3.496&  -7.356\\
1&   5&   0&   -7.32&   -14.174& 2&   1&   0.005&   -35.72&  -74.018& 2.5& 9&   0&   -3.155&  -6.784\\
1&   5.5& 0&   -6.529&  -12.711& 2&   1.5& 0&   -24.617& -49.583& 2.5& 9.5& 0&   -2.842&  -6.266\\
1&   6&   0&   -5.852&  -11.478& 2&   2&   0&   -18.793& -37.182& 2.5& 10&  0&   -2.554&  -5.794\\
1&   6.5& 0&   -5.264&  -10.422& 2&   2.5& 0&   -15.166& -29.65&  2.5& 15&  0&   -0.504&  -2.611\\
1&   7&   0&   -4.746&  -9.507&  2&   3&   0&   -12.666& -24.572& 2.5& 20&  0&   0.774&  -0.799\\
1&   7.5& 0&   -4.285&  -8.704&  2&   3.5& 0&   -10.824& -20.906&2.5& 25&  0&   1.698&   0.429\\
1&   8&   0&   -3.871&  -7.992&  2&   4&   0&   -9.399&  -18.126& 2.5& 30&  0&   2.417&   1.343\\
\end{tabular}
\end{center}
\newpage
\begin{center}
\begin{tabular}{ccccc|ccccc|ccccc} \\ \hline
$log\rho Y_{e}$& $T_{9}$ & $E_{f}$ & $\lambda^{\nu}$ &
$\lambda^{\bar{\nu}}$ & $log\rho Y_{e}$& $T_{9}$ & $E_{f}$ &
$\lambda^{\nu}$ & $\lambda^{\bar{\nu}}$ & $log\rho Y_{e}$& $T_{9}$
& $E_{f}$ & $\lambda^{\nu}$ & $\lambda^{\bar{\nu}}$   \\\hline
3&   0.5& 0.311&   -64.406& -100&    4&  0.5& 0.411&   -63.394& -100&    5&   0.5& 0.522&   -62.276& -100\\
3&   1&   0.046&   -35.514& -74.224& 4&   1&   0.209&   -34.692& -75.046& 5&   1&   0.413&   -33.661& -76.078\\
3&   1.5& 0.005&   -24.602& -49.597& 4&   1.5& 0.047&   -24.461&-49.739& 5&   1.5& 0.265&   -23.727& -50.473\\
3&   2&   0.001&    -18.79&  -37.185& 4&   2&   0.014&   -18.759& -37.216& 5&   2&   0.128&   -18.47&  -37.505\\
3&   2.5& 0.001&    -15.165& -29.651& 4&   2.5& 0.006&   -15.153& -29.662& 5&   2.5& 0.062&   -15.041& -29.775\\
3&  3&   0&   -12.665& -24.572& 4&   3&   0.004&   -12.66&  -24.578& 5&   3&   0.035&   -12.606& -24.631\\
3&   3.5& 0&   -10.823& -20.906& 4&   3.5& 0.002&   -10.82&  -20.909& 5&   3.5& 0.023&  -10.791& -20.938\\
3&   4&   0&   -9.399&  -18.126& 4&   4&   0.002&   -9.397&  -18.128& 5&   4&   0.016&   -9.379&  -18.146\\
3&   4.5& 0&   -8.258&  -15.941& 4&   4.5& 0.001&    -8.257&  -15.942& 5&   4.5& 0.012&   -8.245&  -15.954\\
3&   5&   0&   -7.319&  -14.173& 4&   5&   0.001&    -7.318&  -14.174& 5&   5&   0.009&   -7.31&   -14.182\\
3&  5.5& 0&   -6.528&  -12.71&  4&   5.5& 0.001&    -6.528&  -12.711& 5&   5.5& 0.007&   -6.522&  -12.717\\
3&   6&   0&   -5.851&  -11.477& 4&   6&   0.001&    -5.851&  -11.478& 5&   6&   0.006&   -5.846&  -11.482\\
3&   6.5& 0&   -5.263&  -10.422& 4&   6.5& 0.001&    -5.262&  -10.422& 5&   6.5& 0.005&   -5.259&  -10.425\\
3&  7&   0&   -4.745&  -9.506&  4&   7&   0&   -4.744&  -9.506&  5&   7&   0.004&   -4.742&  -9.509\\
3&   7.5& 0&   -4.284&  -8.703&  4&   7.5& 0&   -4.284&  -8.703&  5&   7.5& 0.004&   -4.281&  -8.705\\
3&   8&   0&   -3.87&  -7.991&  4&   8&   0&   -3.87&   -7.991&  5&   8&   0.003&   -3.868&  -7.993\\
3&   8.5& 0&   -3.496&  -7.356&  4&   8.5& 0&   -3.496&  -7.356&  5&   8.5& 0.003&   -3.494&  -7.357\\
3&   9&   0&   -3.155&  -6.784&  4&   9&   0&   -3.155&  -6.784&  5&   9&   0.002&   -3.153&  -6.785\\
3&   9.5& 0&   -2.842&  -6.266&  4&   9.5& 0&   -2.842&  -6.266&  5&   9.5& 0.002&   -2.841&  -6.267\\
3&   10&  0&   -2.554&  -5.794&  4&   10&  0&   -2.554&  -5.794&  5&   10&  0.002&   -2.553&  -5.794\\
3&   15&  0&   -0.503&  -2.611&  4&   15&  0&   -0.503&  -2.61 &  5&   15&  0.001&    -0.503&  -2.611\\
3&   20&  0&   0.775&   -0.799&  4&   20&  0 &  0.775&   -0.799&  5&   20&  0&   0.775&   -0.799\\
3&   25&  0&   1.699&   0.429&   4&   25&  0&   1.699&   0.429 &  5&   25&  0&   1.699&   0.429\\
3&   30&  0&   2.417&   1.343&   4&   30&  0&   2.417&   1.343&   5&   30&  0&   2.417&   1.343\\

3.5& 0.5& 0.361&   -63.904& -100&    4.5& 0.5& 0.464&   -62.865& -100&    5.5& 0.5& 0.598&   -61.518& -100\\
3.5& 1&   0.115&   -35.167& -74.571& 4.5& 1&   0.309&   -34.187& -75.552& 5.5& 1&   0.528&   -33.082& -76.657\\
3.5& 1.5& 0.015&   -24.568&-49.632& 4.5& 1.5& 0.13&    -24.182& -50.017& 5.5& 1.5& 0.423&   -23.198& -51.003\\
3.5& 2&   0.004&   -18.783& -37.192& 4.5& 2&   0.044&   -18.683& -37.292&5.5& 2&   0.295&   -18.049& -37.926\\
3.5& 2.5& 0.002&   -15.162& -29.653& 4.5& 2.5& 0.02&    -15.126& -29.689& 5.5& 2.5& 0.18&    -14.803& -30.013\\
3.5& 3&   0.001&    -12.664& -24.574& 4.5& 3&   0.011&   -12.647& -24.591& 5.5& 3&   0.11&    -12.482& -24.756\\
3.5& 3.5& 0.001&    -10.822& -20.907& 4.5& 3.5& 0.007&   -10.813& -20.916& 5.5& 3.5& 0.072&   -10.72&  -21.009\\
3.5& 4&   0.001&   -9.398&  -18.127& 4.5& 4 &  0.005&   -9.393&  -18.132& 5.5& 4&   0.05&    -9.336&  -18.19\\
3.5& 4.5& 0&   -8.258&  -15.941& 4.5& 4.5& 0.004&   -8.254&  -15.945& 5.5& 4.5& 0.038&   -8.216&  -15.983\\
3.5& 5&   0&   -7.319&  -14.173& 4.5& 5&   0.003&   -7.316&  -14.176& 5.5& 5&   0.029&   -7.29&   -14.202\\
3.5& 5.5& 0&   -6.528&  -12.71&  4.5& 5.5& 0.002&   -6.526&  -12.712& 5.5& 5.5& 0.023&   -6.507&  -12.731\\
3.5& 6&   0&   -5.851& -11.477& 4.5& 6&   0.002&   -5.85&   -11.479& 5.5& 6&   0.019&   -5.835&  -11.493\\
3.5& 6.5& 0&   -5.263&  -10.422& 4.5& 6.5& 0.002&   -5.261&  -10.423& 5.5& 6.5& 0.016&   -5.25&   -10.434\\
3.5& 7&   0&   -4.745&  -9.506&  4.5& 7&   0.001&    -4.744&  -9.507&  5.5& 7&   0.013&   -4.735&  -9.515\\
3.5& 7.5& 0&   -4.284&  -8.703&  4.5& 7.5& 0.001&   -4.283&  -8.703&  5.5& 7.5& 0.012&   -4.276&  -8.71\\
3.5& 8&   0&   -3.87&   -7.991&  4.5& 8&   0.001&    -3.87&   -7.992&  5.5& 8&   0.01&    -3.864&  -7.998\\
3.5& 8.5& 0&   -3.496&  -7.356&  4.5&8.5& 0.001&    -3.495&  -7.356&  5.5& 8.5& 0.009&   -3.491&  -7.361\\
3.5& 9&   0&   -3.155& -6.784&  4.5& 9&   0.001&    -3.154&  -6.784&  5.5& 9&   0.008&   -3.151&  -6.788\\
3.5& 9.5& 0&   -2.842&  -6.266& 4.5& 9.5& 0.001&    -2.842&  -6.266&  5.5& 9.5& 0.007&   -2.838&  -6.269\\
3.5& 10&  0&   -2.554&  -5.794&  4.5& 10&  0.001&    -2.553&  -5.794&  5.5& 10&  0.006&   -2.55&   -5.797\\
3.5& 15&  0&   -0.503&  -2.61&   4.5& 15&  0&   -0.503&  -2.61&   5.5& 15&  0.003&   -0.502&  -2.611\\
3.5& 20&  0&   0.775&   -0.799&  4.5& 20&  0&  0.775&   -0.799&  5.5& 20&  0.001&    0.775&   -0.799\\
3.5& 25&  0&   1.699&  0.429&   4.5& 25&  0&   1.699&   0.429&   5.5& 25&  0.001&    1.699&   0.429\\
3.5& 30&  0&   2.417&   1.343&   4.5& 30&  0&   2.417&   1.343&   5.5& 30&  0.001&    2.418&   1.343\\
\end{tabular}
\end{center}
\newpage
\begin{center}
\begin{tabular}{ccccc|ccccc|ccccc} \\ \hline
$log\rho Y_{e}$& $T_{9}$ & $E_{f}$ & $\lambda^{\nu}$ &
$\lambda^{\bar{\nu}}$ & $log\rho Y_{e}$& $T_{9}$ & $E_{f}$ &
$\lambda^{\nu}$ & $\lambda^{\bar{\nu}}$ & $log\rho Y_{e}$& $T_{9}$
& $E_{f}$ & $\lambda^{\nu}$ & $\lambda^{\bar{\nu}}$   \\\hline
6&   0.5& 0.713&   -60.357& -100&    7&  0.5& 1.217&   -56.685& -100&    8&   0.5& 2.444&   -46.023& -100\\
6&   1&   0.672&   -32.361& -77.381& 7&   1&   1.2& -29.976& -80.038& 8&   1&   2.437&   -24.4&   -85.748\\
6&   1.5& 0.604&   -22.592& -51.61&  7&   1.5& 1.173&   -20.755& -53.519& 8&   1.5& 2.424&   -16.885& -57.677\\
6&   2&   0.512&   -17.504& -38.473& 7&   2&   1.133&   -15.966& -40.037& 8&   2&   2.406&   -12.964& -43.235\\
6&   2.5& 0.405&   -14.351& -30.465& 7&   2.5& 1.083&   -12.997& -31.832& 8&   2.5& 2.383&   -10.511& -34.45\\
6&   3&   0.299&   -12.164& -25.074& 7&   3&   1.021&   -10.959& -26.287& 8&  3&   2.355&   -8.812&  -28.526\\
6&   3.5& 0.214&   -10.515& -21.214& 7&   3.5& 0.95&    -9.461&  -22.274& 8&   3.5& 2.322&   -7.553&  -24.248\\
6&   4&  0.156&   -9.203&  -18.323& 7&   4&   0.871&   -8.307&  -19.223& 8&   4&   2.283&   -6.577&  -21.003\\
6&   4.5& 0.117&   -8.127&  -16.072& 7&   4.5& 0.785&   -7.382&  -16.82&  8&   4.5& 2.24&    -5.792&  -18.449\\
6&   5&   0.091&   -7.227&  -14.265& 7&  5 &  0.698&   -6.618&  -14.877& 8&   5 &  2.192&   -5.144&  -16.382\\
6&   5.5& 0.073&   -6.461&  -12.777& 7&   5.5& 0.613&   -5.969&  -13.272& 8&   5.5& 2.139&   -4.597&  -14.67\\
6&   6&   0.06&   -5.801&  -11.527& 7&   6 &  0.534&   -5.404&  -11.926& 8&   6&   2.081&   -4.127&  -13.225\\
6&   6.5& 0.05 &   -5.224&  -10.46&  7&   6.5& 0.465&   -4.904&  -10.782& 8&   6.5& 2.019&   -3.718&  -11.987\\
6&   7&   0.042&   -4.714&  -9.536&  7&   7&   0.404&   -4.455&  -9.797&  8&   7&   1.952&   -3.357&  -10.911\\
6&   7.5& 0.036&   -4.259&  -8.727&  7&   7.5& 0.353&   -4.048&  -8.94&   8&   7.5& 1.882&   -3.034&  -9.967\\
6&   8&   0.032&   -3.85&   -8.011&  7&   8&   0.31&    -3.676&  -8.187&  8&   8&   1.808&   -2.744&  -9.13\\
6&   8.5& 0.028&   -3.479&  -7.372&  7&   8.5& 0.274&   -3.334&  -7.518&  8&   8.5& 1.732&   -2.481&  -8.382\\
6&   9&   0.025&   -3.141&  -6.798&  7&   9&   0.244&   -3.019&  -6.92&   8&   9&   1.653&   -2.239&  -7.71\\
6&   9.5& 0.022&   -2.83&   -6.277&  7&   9.5& 0.218&   -2.727&  -6.381&  8&   9.5& 1.574&   -2.016&  -7.1\\
6&   10&  0.02&    -2.544&  -5.803&  7&   10&  0.196&   -2.455&  -5.892&  8&   10&  1.493&   -1.809&  -6.546\\
6&   15&  0.009&   -0.5&    -2.613&  7&   15&  0.085&   -0.475&  -2.639&  8&   15&  0.817&   -0.231&  -2.885\\
6&   20&  0.005&   0.776&   -0.8&    7&   20&  0.047&   0.787&   -0.81&   8&   20&  0.47 &   0.892&   -0.917\\
6&   25&  0.003&   1.699&   0.429&   7&   25&  0.03&    1.705&   0.423&   8&   25&  0.301&   1.759&   0.369\\
6&   30&  0.002&   2.418&   1.343&   7&   30&  0.021&   2.421&   1.34&    8&   30&  0.209&   2.452&   1.309\\

6.5& 0.5& 0.905&   -58.556& -100&    7.5& 0.5& 1.705&   -53.469& -100&    8.5& 0.5& 3.547&   -34.902& -100\\
6.5& 1&   0.88&    -31.345& -78.427& 7.5& 1&   1.693&   -28.044&-82.349& 8.5& 1&   3.542&   -18.828& -91.319\\
6.5& 1.5& 0.837&   -21.819& -52.393& 7.5& 1.5& 1.675&   -19.269& -55.187& 8.5& 1.5& 3.534&   -13.164& -61.405\\
6.5& 2&   0.777&   -16.842& -39.139& 7.5& 2&   1.648&   -14.756& -41.33&  8.5& 2&   3.521&   -10.181& -46.046\\
6.5& 2.5& 0.701&   -13.756& -31.062& 7.5& 2.5& 1.614&   -11.971& -32.902& 8.5& 2.5& 3.506&   -8.295&  -36.714\\
6.5& 3&   0.612&   -11.64&  -25.6&   7.5& 3&   1.573&   -10.058& -27.214& 8.5& 3&   3.487&   -6.971&  -30.428\\
6.5& 3.5& 0.517&   -10.081& -21.65&  7.5& 3.5& 1.524&   -8.652&  -23.1&   8.5& 3.5& 3.464&   -5.973&  -25.893\\
6.5& 4&   0.424&  -8.866&  -18.66&  7.5& 4&   1.468&   -7.566&  -19.976& 8.5& 4&   3.438 &  -5.185&  -22.457\\
6.5& 4.5& 0.343&   -7.875&  -16.324& 7.5& 4.5& 1.405&   -6.697&  -17.515& 8.5& 4.5& 3.408 &  -4.542&  -19.758\\
6.5& 5&   0.277&  -7.04&   -14.452& 7.5& 5&   1.336&   -5.982&  -15.52&  8.5& 5&   3.375 &  -4.005&  -17.575\\
6.5& 5.5& 0.226&   -6.322 & -12.917& 7.5& 5.5& 1.262&   -5.381&  -13.866& 8.5& 5.5& 3.339&   -3.546&  -15.769\\
6.5& 6&   0.187&   -5.695 & -11.634& 7.5& 6&   1.183&   -4.865&  -12.47&  8.5& 6&   3.299&   -3.147&  -14.248\\
6.5& 6.5& 0.157&   -5.141&  -10.543& 7.5& 6.5& 1.101&   -4.415&  -11.275& 8.5& 6.5& 3.256&   -2.798&  -12.945\\
6.5& 7&   0.134&   -4.649&  -9.602&  7.5& 7&   1.018&   -4.016&  -10.239& 8.5& 7&   3.209&   -2.487&  -11.816\\
6.5& 7.5& 0.115&   -4.207&  -8.78&   7.5& 7.5& 0.937&   -3.659&  -9.332&  8.5& 7.5& 3.159&   -2.208&  -10.825\\
6.5& 8&   0.1& -3.807&  -8.054&  7.5& 8&   0.858&   -3.333&  -8.532&  8.5& 8&   3.106&   -1.955&  -9.948\\
6.5& 8.5& 0.088&   -3.444&  -7.408&  7.5& 8.5& 0.783&   -3.035&  -7.82&   8.5& 8.5& 3.05&    -1.725&  -9.164\\
6.5& 9&   0.078&   -3.111&  -6.827&  7.5& 9&   0.714&   -2.758&  -7.184&  8.5& 9 &  2.99&    -1.514&  -8.458\\
6.5& 9.5& 0.069&   -2.805&  -6.302&  7.5& 9.5& 0.651&   -2.499&  -6.611&  8.5& 9.5& 2.928&   -1.319&  -7.819\\
6.5& 10&  0.062&  -2.522&  -5.825&  7.5& 10&  0.593&   -2.257&  -6.092&  8.5& 10&  2.863&   -1.138&  -7.236\\
6.5& 15&  0.027&   -0.494&  -2.619&  7.5& 15&  0.268&   -0.414&  -2.7&    8.5& 15&  2.109&   0.195&   -3.318\\
6.5& 20&  0.015&  0.779&   -0.802&  7.5& 20&  0.15&    0.812&   -0.836&  8.5& 20& 1.402&   1.124&   -1.152\\
6.5& 25&  0.01 &   1.701&   0.428&   7.5& 25&  0.095&   1.718&   0.41&    8.5& 25&  0.936&   1.885&   0.241\\
6.5&30&  0.007&   2.419&   1.342&   7.5& 30&  0.066&   2.428&   1.332&   8.5& 30&  0.656&   2.527&   1.233\\
\end{tabular}
\end{center}
\newpage
\begin{center}
\begin{tabular}{ccccc|ccccc|ccccc} \\ \hline
$log\rho Y_{e}$& $T_{9}$ & $E_{f}$ & $\lambda^{\nu}$ &
$\lambda^{\bar{\nu}}$ & $log\rho Y_{e}$& $T_{9}$ & $E_{f}$ &
$\lambda^{\nu}$ & $\lambda^{\bar{\nu}}$ & $log\rho Y_{e}$& $T_{9}$
& $E_{f}$ & $\lambda^{\nu}$ & $\lambda^{\bar{\nu}}$   \\\hline
9&   0.5&  5.179&    -18.453&  -100&     9.5&  8.5&  7.351&    0.661&    -11.712&  10.5&     4.5&  16.28&    3.848&    -34.166\\
9&    1&    5.176&    -10.596&  -99.551&  9.5&  9&    7.323&    0.749&    -10.883&  10.5&     5&    16.273&   3.848&    -30.569\\
9&    1.5&  5.17&     -7.666&   -66.903&  9.5&  9.5&  7.293&    0.836&    -10.133&  10.5&     5.5&  16.265&   3.847&    -27.609\\
9&    2&    5.162&    -6.048&   -50.178&  9.5& 10&   7.261&    0.922&    -9.452&   10.5&     6&    16.256&   3.846 &   -25.127\\
9&    2.5&  5.151&    -4.983&   -40.029&  9.5&  15&   6.86&     1.689&    -4.914&   10.5&     6.5&  16.247&   3.845&    -23.014\\
9&    3&    5.138&    -4.208&   -33.2&    9.5&  20&   6.307&    2.304&    -2.387&   10.5&     7&    16.237&   3.845&    -21.192\\
9&    3.5&  5.122&    -3.607&   -28.279&  9.5&  25&   5.624&    2.8&  -0.704&   10.5&     7.5&  16.226&   3.845 &   -19.603\\
9&    4&    5.105&    -3.119&   -24.555&  9.5&  30&   4.859&    3.216&    0.528&    10.5&     8&    16.214&   3.846&    -18.203\\
9&    4.5&  5.085&    -2.71&    -21.633&  10&   0.5&  11.118&   2.13&     -100&     10.5&     8.5&  16.202&   3.847&    -16.959\\
9&    5&    5.062&    -2.358&   -19.273&  10&   1&    11.116&   2.133&    -100&     10.5&     9&    16.189&   3.849&    -15.847\\
9&    5.5&  5.037&    -2.049&   -17.324&  10&   1.5&  11.113&   2.139&    -86.813&  10.5&    9.5&  16.175&   3.852&    -14.844\\
9&    6&    5.01&     -1.773&  -15.684&  10&   2&    11.11&    2.146&    -65.139&  10.5 &    10&   16.16 &   3.856&    -13.936\\
9&    6.5&  4.98&     -1.525&   -14.281&  10&   2.5&  11.105&   2.154&    -52.016&  10.5&     15&   15.973&   3.964&    -7.976\\
9&    7&    4.948&    -1.3&     -13.067& 10&   3&    11.099&   2.164&    -43.204&  10.5&     20&   15.711&   4.192&   -4.757\\
9&    7.5&  4.914&    -1.093&   -12.003&  10&   3.5&  11.091&   2.175&    -36.867&  10.5&     25&   15.375&   4.461&   -2.669\\
9&    8&    4.878&    -0.902&   -11.063&  10&   4&    11.083&   2.186&    -32.083&  10.5&     30&   14.965&   4.716&   -1.17\\
9&    8.5&  4.839&    -0.724&   -10.224&  10&   4.5&  11.074&   2.198&    -28.337&  11&   0.5&  23.934&   5.232&    -100\\
9&    9&    4.797&    -0.559&   -9.469&   10&   5&    11.063&   2.21&     -25.319&  11&   1&    23.933&   5.232&   -100\\
9&    9.5&  4.754&    -0.405&   -8.787&   10&   5.5&  11.052&   2.224&    -22.833&  11&   1.5&  23.932&   5.233&    -100\\
9&    10&   4.708&    -0.26&    -8.165&   10&   6&    11.039&  2.238&    -20.746&  11&   2&    23.93&    5.233&    -97.44\\
9&    15&   4.131&    0.848&    -3.998&   10&   6.5&  11.025&   2.254&    -18.967&  11&   2.5&  23.928&   5.233&    -77.863\\
9&    20&   3.39&     1.612&    -1.652&   10&  7&    11.011&  2.27&     -17.43&   11&   3&    23.925&   5.233&    -64.749\\
9&    25&   2.621&    2.219&    -0.098&   10&   7.5&  10.995&   2.288&    -16.088& 11&   3.5&  23.922&   5.231&    -55.34\\
9&    30&   1.973&    2.744&    1.013&    10&   8&    10.978&   2.308&    -14.904&  11&   4&    23.918&   5.229&    -48.252\\
9.5&  0.5&  7.583&    -0.829&   -100&     10&   8.5&  10.959&   2.329&    -13.852&  11&   4.5&  23.913&   5.226&   -42.715\\
9.5&  1&    7.581&    -0.787&   -100&     10&   9&    10.94&    2.351&    -12.908&  11&   5&    23.908&   5.222&    -38.266\\
9.5&  1.5&  7.577&    -0.721&   -74.939&  10&   9.5&  10.92&    2.376&   -12.057& 11&   5.5&  23.903&   5.218&    -34.608\\
9.5&  2&    7.571&    -0.637&  -56.226&  10&   10&   10.898&   2.402&    -11.284&  11&   6&    23.897&   5.214&   -31.545\\
9.5&  2.5&  7.564&    -0.541&   -44.88&   10&   15&   10.624&   2.75&     -6.179&  11&   6.5&  23.891&   5.209&    -28.941\\
9.5&  3&    7.555&    -0.437&   -37.252&  10&   20&   10.241&   3.167&    -3.378&   11&   7&    23.884&  5.205&    -26.698\\
9.5&  3.5&  7.545&    -0.329&   -31.76&   10&   25&   9.751&    3.554&    -1.536&   11&   7.5&  23.877&   5.2&  -24.744\\
9.5&  4&    7.532&    -0.221&   -27.61&   10&   30&   9.163&    3.891&    -0.195&  11&   8&    23.869&   5.196&    -23.025\\
9.5&  4.5&  7.519&   -0.114&   -24.356&  10.5&     0.5&  16.31&    3.841&    -100&     11&   8.5&  23.86&    5.192&    -21.5\\
9.5&  5&    7.503&    -0.009&   -21.731&  10.5&     1&    16.309&   3.842&    -100&     11&   9&    23.851&   5.188&    -20.138\\
9.5&  5.5&  7.486&    0.093&    -19.566&  10.5&     1.5&  16.307&   3.843&    -100&     11&   9.5&  23.842&   5.185&    -18.912\\
9.5&  6&    7.468&    0.193&    -17.746&  10.5&     2&    16.304&  3.845&    -78.224&  11&   10&   23.832&   5.182 &   -17.802\\
9.5&  6.5&  7.448&    0.291&    -16.193&  10.5&     2.5&  16.301&   3.846&    -62.488&  11&   15&   23.704&   5.196&    -10.574\\
9.5&  7&    7.426&    0.386&    -14.849&  10.5&     3&    16.297&   3.848&   -51.934&  11&   20&   23.526&   5.309&    -6.726\\
9.5&  7.5&  7.403&    0.479&    -13.674&  10.5&     3.5&  16.292&   3.848&    -44.354&  11&   25 &  23.296&   5.484&    -4.266\\
9.5&  8&    7.378&    0.571&    -12.637&  10.5&     4&    16.286&
3.849&   -38.637&  11&   30&   23.016&   5.669&    -2.522\\\hline

\end{tabular}
\end{center}}

\begin{figure}[htbp]
\caption{(Color online) Neutrino energy loss rates due to
$^{24}$Mg, as a function of stellar temperatures, for different
selected stellar densities . Densities are given in units of
$gcm^{-3}$. Temperatures are measured in $10^{9}$ K and
log$\lambda_{\nu}$ represents the log (to base 10) of neutrino
energy loss rates in units of $MeV sec^{-1}$.}
\begin{center}
\begin{tabular}{c}
\includegraphics[width=0.8\textwidth]{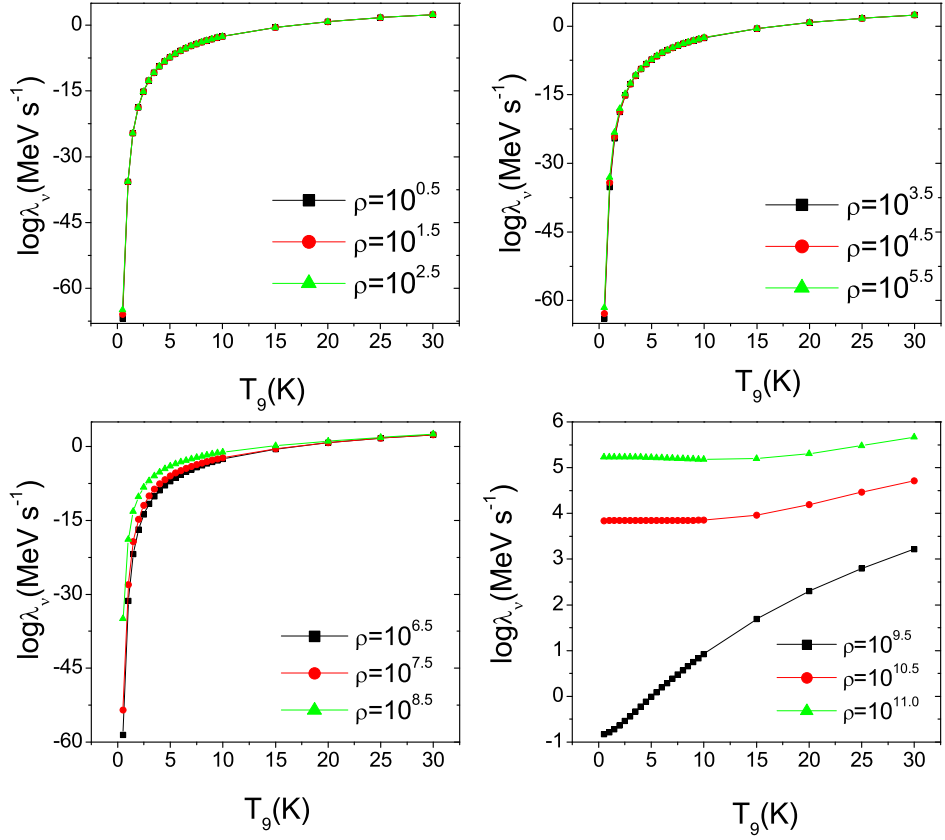}
\end{tabular}
\end{center}
\end{figure}

\begin{figure}[htbp]
\caption{(Color online) Same as Fig.1 but for antineutrino energy
loss rates.}
\begin{center}
\begin{tabular}{c}
\includegraphics[width=0.8\textwidth]{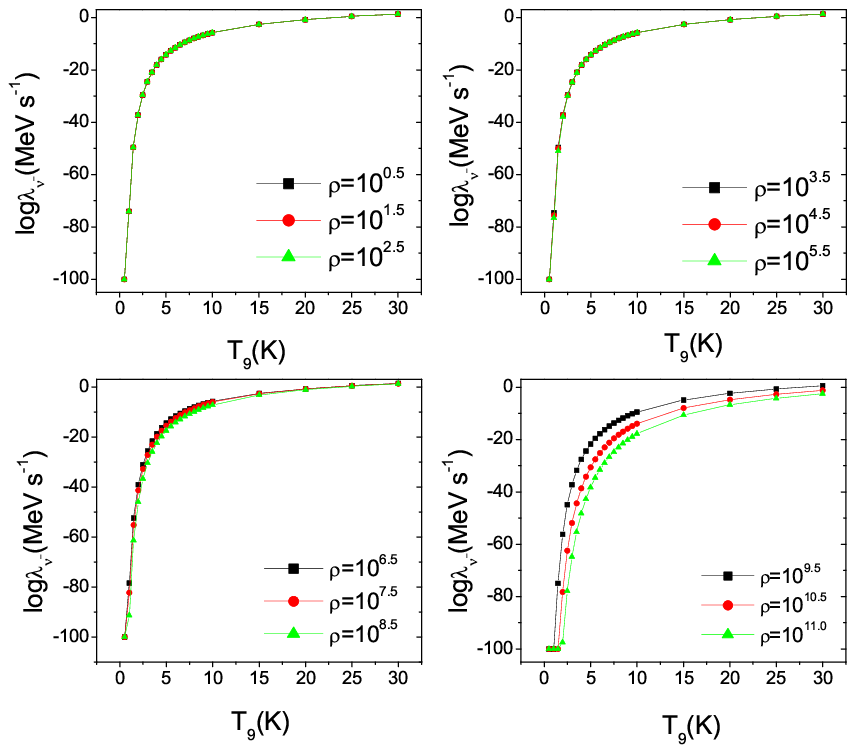}
\end{tabular}
\end{center}
\end{figure}

\begin{figure}[htbp]
\caption{(Color online) Comparison of pn-QRPA calculated neutrino
energy loss rates with those calculated using shell model (OHMTS)
[26] and those calculated by FFN [36] as a function of stellar
temperatures and densities. Ordinate represents the log (to base
10) of neutrino energy loss rates in units of $MeV sec^{-1}$.}
\begin{center}
\begin{tabular}{c}
\includegraphics[width=1.1\textwidth]{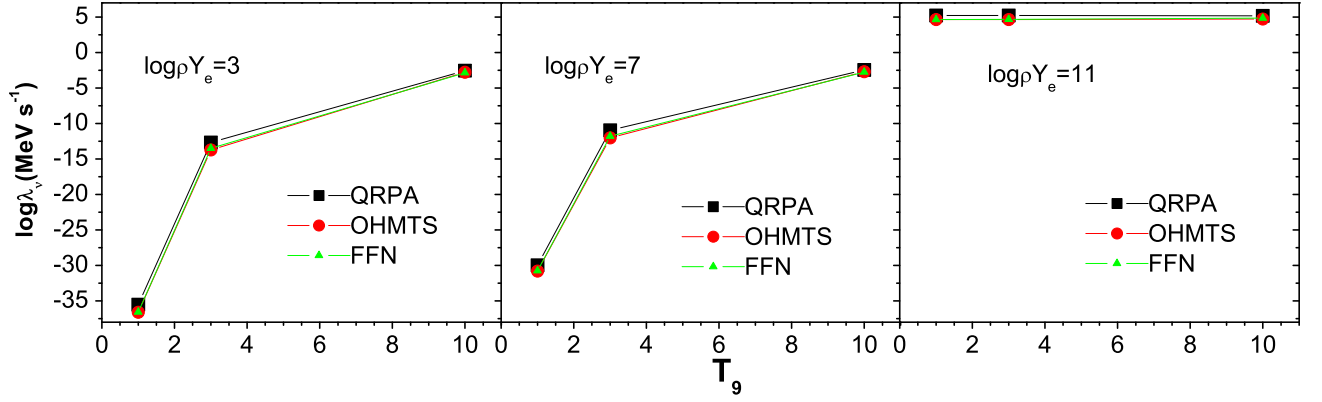}
\end{tabular}
\end{center}
\end{figure}

\begin{figure}[htbp]
\caption{(Color online) Same as Fig.3 but for antineutrino energy
loss rates.}
\begin{center}
\begin{tabular}{c}
\includegraphics[width=1.1\textwidth]{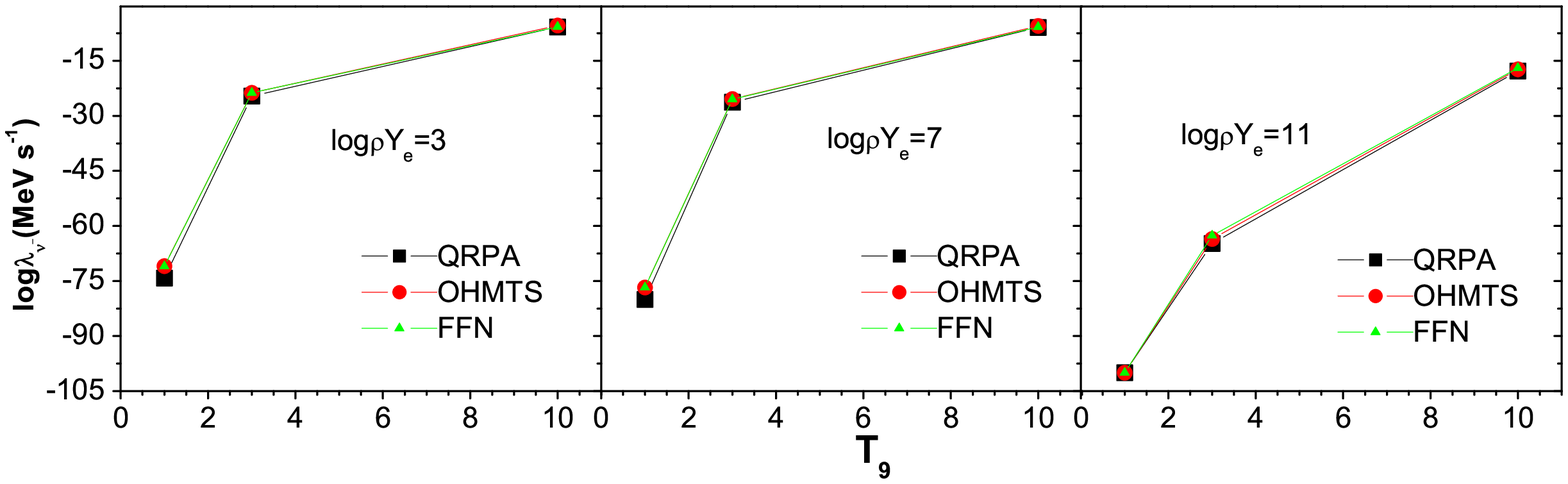}
\end{tabular}
\end{center}
\end{figure}

\end{document}